\documentclass[twocolumn,aps,superscriptaddress,showpacs,floatfix]{revtex4}
\usepackage{graphicx}
\usepackage{booktabs}
\usepackage{amssymb,bm,mathrsfs,bbm,amscd}
\usepackage[tbtags]{amsmath}
\usepackage{lastpage}
\usepackage{CJK}\usepackage{epsfig}

\begin{document}
\begin{CJK*}{GBK}{song}

\title{Single particle potentials of asymmetric nuclear matter in different spin-isospin channels}

\author{Wei Zuo\footnote{zuowei@impcas.ac.cn}}
\affiliation{Institute of Modern Physics, Chinese Academy of
Sciences, Lanzhou 730000, China}
 \author{ Sheng-Xin Gan}\affiliation{Institute of Modern Physics, Chinese Academy of
Sciences, Lanzhou 730000, China} \affiliation{University
of Chinese Academy of Sciences, Beijing, 100049, China}
\author{Umberto Lombardo}\affiliation{Department of Physics and Astrophysics, Catania University,
Via Santa Sofia 64, I-95123, Italy}


\begin{abstract}
We investigate the neutron and proton single particle (s.p.)
potentials of asymmetric nuclear matter and their isospin dependence
in various spin-isospin $ST$ channels within the framework of the
Brueckner-Hartree-Fock approach. It is shown that in symmetric
nuclear matter, the s.p. potentials in both the isospin-singlet $T=0$
channel and isospin-triplet $T=1$ channel are essentially
attractive, and the magnitudes in the two different channels are
roughly the same. In neutron-rich nuclear matter, the
isospin-splitting of the proton and neutron s.p. potentials turns
out to be mainly determined by the isospin-singlet $T=0$ channel
contribution which becomes more attractive for proton and more
repulsive for neutron at higher asymmetries.
\\[4mm]
{\bf keyword:} single particle potential, asymmetric nuclear matter,
 contribution from various spin-isospin channels, symmetry potential,
 Brueckner-Hartree-Fock approach
\end{abstract}

\maketitle

\section{Introduction}

 One of the main aims of nuclear physics
 is to determine the equation of state (EOS) of nuclear matter and
 constrain effective nucleon-nucleon (NN)
interactions. Nucleon single particle (s.p.) potentials, especially their
isospin dependence in asymmetric
 nuclear matter are basic inputs for the dynamic simulations of
 heavy ion collisions and are expected to play an important role in constraining
nucleon-nucleon (NN) effective
interactions in asymmetric nuclear matter\cite{li:2008}.
The phenomenological NN effective interactions such as the Skyrme and Skyrme-like interactions play an
important role in predicting the properties of finite
nuclei~\cite{vautherin:1972,friedrich:1986,dobaczewski:1996,%
goriely:2002,goriely:2003,lesinski:2007,brito:2007}, nuclear matter
and neutron
stars~\cite{onsi:2002,stone:2003,stone:2006,meissner:2007},
nucleus-nucleus interaction potential~\cite{denisov:2002,wang:2006}
and fission barriers~\cite{goriely:2007}. The parameters of the
effective interactions are usually constrained by the ground state
properties of stable nuclei and/or the saturation properties of nuclear
matter, and thus they are shown to be quite successful for
describing nuclear phenomena related to nuclear system not far from
the normal nuclear matter density ($\rho_0=0.17$fm$^{-3}$) at small
isospin-asymmetries. However, as soon as the density deviates from
the normal nuclear matter density and the isospin-asymmetry becomes
large, the discrepancy among the predictions of the
Skyrme-Hartree-Fock (SHF) approach by adopting different Skyrme
parameters could be extremely large~\cite{babrown:2000,chen:2005}.
As for the isospin dependence of nucleon single-particle properties,
different Skyrme parameters may lead to an opposite isospin
splitting of the neutron and proton effective masses in neutron-rich
nuclear matter even at densities around
$\rho_0$~\cite{lesinski:2006}. In order to improve the predictive
power of the Skyrme interaction at high densities and large isospin
asymmetries, some work was done in recent years to constrain
the Skyrme parameters by fitting the bulk properties of asymmetric
nuclear matter obtained by the SHF approach to those predicted by
microscopic many-body theories. For example, in Ref.~\cite{chabanat}
Chabanat {\it et al.} proposed a number of sets of Skyrme parameters
by reproducing the equation of states (EOSs) of symmetric nuclear
matter and pure neutron matter predicted by the microscopic
variational approach~\cite{pudliner:1995}. In Ref.~\cite{cao:2006},
the authors constructed the LNS parameters for the Skyrme
interaction by fitting to the EOS of asymmetric nuclear matter and
the neutron/proton effective mass splitting in neutron-rich matter
around saturation density obtained within the Brueckner-Hartree-Fock
 (BHF) approach extended to include a microscopic three-body force
(TBF)~\cite{zuo:2002,zuo:2005}. Although these recent
parametrizations of Skyrme interaction can reproduce fairly well the
EOSs of symmetric nuclear matter and pure neutron matter predicted
by microscopic approaches(variational method and BHF approach), the
deviation from the microscopic results is shown to become
significantly large even for symmetric nuclear matter as soon as the
EOS is decomposed into different spin-isospin
channels~\cite{lesinski:2006}, and what is more,
the single particle (s.p.) properties (especially their isospin and momentum dependence)
obtained by the Skyrme-Hartree-Fock calculation could be significantly different
from the predictions of microscopic approaches.

In Ref.~\cite{zuo:2009}, we investigated the EOS of asymmetric
nuclear matter and its isospin dependence in various spin-isospin
$ST$ channels within the framework of the microscopic BHF approach.
In the present paper, we shall extend our work to study the proton
and neutron s.p. potentials in different spin-isospin channels for a
deeper understanding of the mechanism of the isospin dependence of
the nuclear matter properties and for providing more elaborate microscopic
constraints for effective \mbox{NN} interactions. We shall discuss
particularly the isovector part and the isospin dependence of the
neutron and proton s.p. potentials in asymmetric nuclear matter in different spin-isospin $ST$
channels.

\section{Theoretical Approaches}

Our present investigation is based on the Brueckner theory~\cite{day:1967}.
The Brueckner approach for asymmetric
nuclear matter and its extension to include a microscopic TBF can be
found in Refs.~\cite{zuo:2002,zuo:1999}. Here we simply give a brief
review for completeness. The starting point of the BHF approach is
the reaction $G$-matrix, which satisfies the following isospin
dependent Bethe-Goldstone (BG) equation,
\begin{eqnarray}
G(\rho, \beta, \omega )&=& \upsilon_{NN} +\upsilon_{NN}
 \nonumber \\ &\times&
\sum_{k_{1}k_{2}}\frac{ |k_{1}k_{2}\rangle Q(k_{1},k_{2})\langle
k_{1}k_{2}|}{\omega -\epsilon (k_{1})-\epsilon (k_{2})}G(\rho,
\beta, \omega ) ,
\end{eqnarray}
where $k_i\equiv(\vec k_i,\sigma_1,\tau_i)$, denotes the momentum,
the $z$-component of spin and isospin of a nucleon, respectively.
$\upsilon_{NN}$ is the realistic NN interaction, and $\omega$ is the
starting energy. The asymmetry parameter is defined as
$\beta=(\rho_n-\rho_p)/\rho$, where $\rho, \rho_n$, and $\rho_p$
denote the total, neutron and proton number densities, respectively.
In solving the BG equation for the $G$-matrix, the continuous
choice~\cite{jeukenne:1976} for the auxiliary potential $U(k)$ is
adopted since it provides a much faster convergence of the hole-line
expansion than the gap choice~\cite{song:1998}. Under the continuous
choice, the auxiliary potential describes the BHF mean field felt by
a nucleon during its propagation in nuclear
medium~\cite{lejeune:1978}.

The BG equation has been solved in the total angular momentum
representation~\cite{zuo:1999}. By using the standard
angular-averaging scheme for the Pauli operator and the energy
denominator, the BG equation can be decoupled into different partial
wave $\alpha=\{JST\}$ channels~\cite{baldo:1999}, where $J$ denotes
the total angular momentum, $S$ the total spin and $T$ the total
isospin of a two-particle state.

For the NN interaction, we adopt the Argonne $V_{18}$ ($AV_{18}$)
two-body interaction~\cite{wiringa:1995} plus a microscopic based on
the meson-exchange current approach~\cite{grange:1989}. The
parameters of the TBF model have been self-consistently determined
so as to reproduce the $AV_{18}$ two-body force by using the
one-boson-exchange potential model~\cite{zuo:2002}. The TBF contains
the contributions from different intermediate virtual processes such
as virtual nucleon-antinucleon pair excitations, and nucleon
resonances ( for details, see Ref.~\cite{grange:1989}). The TBF
effects on the EOS of nuclear matter and its connection to the
relativistic effects in the DBHF approach have been reported in
Ref.~\cite{zuo:2002}.

The TBF contribution has been included by reducing the TBF to an
equivalently effective two-body interaction via a suitable average
with respect to the third-nucleon degrees of freedom according to
the standard scheme~\cite{grange:1989}. The effective two-body
interaction ${\tilde v}$ can be expressed in $r$-space
as\cite{zuo:2002}
\begin{equation}
\begin{array}{lll}
 \langle\vec r_1 \vec r_2| {\tilde v} |
\vec r_1^{\ \prime} \vec r_2^{\ \prime} \rangle = \displaystyle
\frac{1}{4} {\rm Tr}\sum_n \int {\rm d} {\vec r_3} {\rm d} {\vec
r_3^{\ \prime}}\phi^*_n(\vec r_3^{\ \prime})(1-\eta(r_{23}'))
 \\[5mm]\displaystyle
 \times
 \displaystyle (1-\eta(r_{13}' ))W_3(\vec r_1^{\ \prime}\vec r_2^{\
\prime} \vec r_3^{\ \prime}|\vec r_1 \vec r_2 \vec r_3)
 (1-\eta(r_{13}))\\[3mm] \times (1-\eta(r_{23})) \phi_n(r_3)
\end{array}\label{eq:TBF}
\end{equation}
where the trace is taken with respect to the spin and isospin of the
third nucleon. The function $\eta(r)$ is the defect function. Since
the defect function is directly determined by the solution of the BG
equation\cite{grange:1989}, it must be calculated self-consistently
with the $G$ matrix and the s.p. potential $U(k)$\cite{zuo:2002} at
each density and isospin asymmetry. It is evident from
Eq.(\ref{eq:TBF}) that the effective force ${\tilde v}$ rising from
the TBF in nuclear medium is density dependent. A detailed
description and justification of the method can be found in
Ref.~\cite{grange:1989}.

\section{Results and Discussion}
\begin{center}
\begin{figure}\includegraphics[width=8cm]{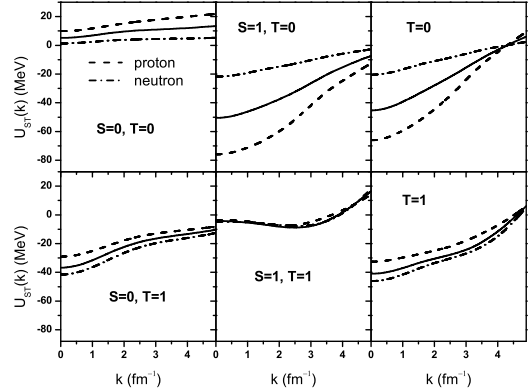}
\caption{
 Decomposition of the neutron (the dash-dotted curves) and proton (the dashed curves)
 s.p. potentials into various spin-isospin $ST$ channels in
 asymmetric nuclear matter at density $\rho=0.17$fm$^{-3}$ and isospin-asymmetry $\beta=0.6$.
 The results for symmetric matter are also shown by the solid curves for comparison.}
\label{fig1}
\end{figure}\end{center}
\begin{center}
\begin{figure}\includegraphics[width=8cm]{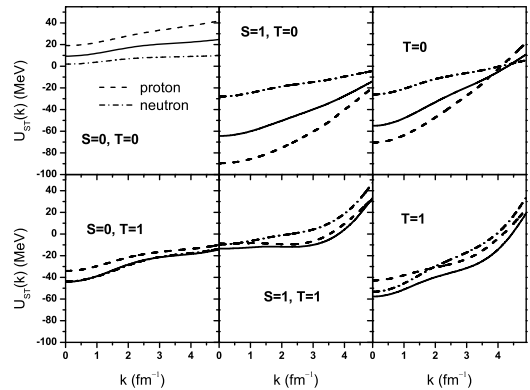}
\caption{
 The same as Fig.~\ref{fig1} but for a density $\rho=0.34$fm$^{-3}$.
 }
\label{fig2}
\end{figure}\end{center}
\begin{center}
\begin{figure}
\includegraphics[width=8cm]{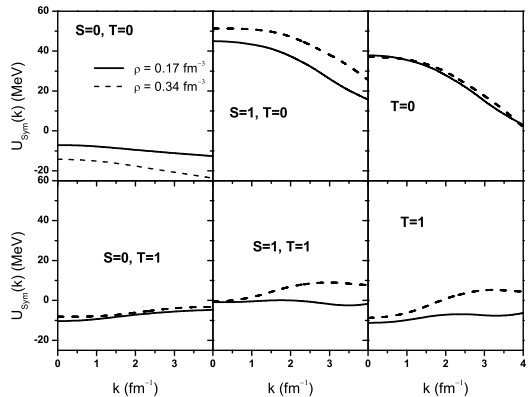}
 \caption{
Decomposition of nuclear symmetry potential into various spin-isospin channels for
 two values of density $\rho=0.17$ and 0.34fm$^{-3}$,
 respectively.
  }
\label{fig3}
\end{figure}
\end{center}

In Fig.~\ref{fig1} we display the neutron (the dash-dotted curves) and
proton (the dashed curves) s.p. potentials in various spin-isospin
channels of $ST =00, 10, 01, 11$, and $T=0, 1$ at a density of
$\rho=0.17$fm$^{(-3)}$ and an isospin-asymmetry of $\beta=0.6$.
Shown in Fig.~\ref{fig2} are the results for
$\rho=0.34$fm$^{-3}$. In the two figures, the s.p. potentials in
symmetric nuclear matter are also plotted (solid lines) for
comparison. It is seen that in symmetric nuclear matter, the s.p.
potentials in the isospin-singlet $T=0$ channel and in the
isospin-triplet $T=1$ channel are attractive and are compatible in
magnitude. One may notice that the contributions in the two even
channels ($ST=10$ and $ST=01$) are considerably larger in magnitude
than those in the two odd channels ($ST=00$ and $ST=11$).
Consequently, the attraction in both the $T=0$ and $T=1$ channels
mainly come from the two even channels. In asymmetric nuclear matter,
the neutron and proton s.p. potentials will split (i.e. become
different) with respect to their common values in symmetric nuclear
matter. It can be seen that the splitting of the proton and neutron
s.p. potentials is much larger in the isospin-singlet $T=0$ channel
than that in the isospin-triplet $T=1$ channel. And thus the
splitting is dominated by the isospin-singlet $T=0$ channel. This
result is consistent with the prediction for the EOS of asymmetric
nuclear matter in~\cite{zuo:1999,zuo:2005} where it is shown that
the isovector part of the EOS of asymmetric nuclear matter is
determined by the contribution from the $T=0$ channel.
As the isospin-asymmetry increases, the proton potential in the $ST=10$ channel
becomes more attractive and the neutron one in the $ST=10$ becomes less attractive.
The isospin-asymmetry dependence of the proton and neutron potentials in the $ST=00$
channel turns out to be opposite
to that in the $ST=10$ channel.
The isospin dependence of the proton and neutron potentials in the
two isospin-triplet ($ST=01$ and $ST=11$) channels is quite weak.
At densities around the nuclear saturation density $\rho=0.17$fm$^{-3}$,
the attraction decreases for the proton s.p. potential and increases for the neutron
s.p. potential slightly in the $T=1$ channel as the asymmetry increases.
At a high density ($\rho=0.34$fm$^{-3}$), the attraction in the $T=1$ channel becomes
slightly smaller for both proton and neutron at a higher asymmetry.
It is also seen from Fig. 1 and Fig.2 that
the isospin dependence of the neutron and proton s.p. potentials in the $T=0$
channel becomes weaker as the momentum increases, which responds for the
repaid decreasing of nuclear symmetry potential as a function of
momentum~\cite{zuo:2005}.
To see more clearly the isospin dependence of the s.p. potentials in asymmetric nuclear matter,
we show in Fig. 3 the contribution to nuclear symmetry potential from different $ST$
channels at two densities $\rho=0.17$fm$^{-3}$ and 0.34fm$^{-3}$.
The symmetry potential is defined as: $U_{\rm sym} = (U_n-U_p)/(2\beta)$.
It describes the isovector part of the neutron and
proton s.p. potentials in neutron-rich nuclear matter and is crucial for predicting the isospin observables in
heavy ion collisions at medium and high energies~\cite{zuo:2005,li:2008}. Symmetry potential is defined as: $U_{\rm
sym}=(U_n-U_p)/2\beta$. It describes the isovector part of the s.p.
potential and is crucial for predicting the isospin observables in
heavy ion collisions at medium and high energies~\cite{li:2008}.
The symmetry potential in the isospin-singlet channel ($T=0$) is shown to be positive,
while that in the isospin-triplet channel $T=1$ is negative.
From Fig.~\ref{fig3}, one may see again that the contribution from the $T=0$ channel is much larger in magnitude
than that from the $T=1$ channel especially at low momenta, which indicates that
the momentum-dependence of the isovector part of nucleon s.p. potential in neutron-rich
nuclear matter is governed by the contribution from the $T=0$ channel. The positive symmetry
potential in the isospin-singlet $T=0$ channel implies that it is repulsive on neutrons and
attractive on protons in the momentum range considered here.
In the $T=0$ channel, the contribution from the odd partial wave $ST=00$ channel is negative,
while the contribution from the even partial
wave $ST=10$ channel is positive. At relatively low momenta, the contribution from the $ST=10$
channel turns out to be much larger in magnitude
than that in the $ST=00$ channel. As a consequence, the contribution from the even partial
wave $ST=10$ channel dominates the symmetry potential in the $T=0$
channel and it determines the total symmetry potential to a large extent. It is also noticed
that the symmetry potential in the $T=0$ channel is almost independent on density, while the
symmetry potential in the $T=1$ channel becomes slightly larger at a higher density.
In the $T=0$ channel, the symmetry potential is a decreasing function of momentum and
the decreasing rate is almost completely determined by the contribution from the even channel $ST=10$.
In symmetric nuclear matter, the present investigation shows that the contribution from the isospin-singlet $T=0$ channel is
almost the same as the contribution from the
isospin-triplet $T=1$ channel, indicating that both the interactions between the like
nucleons (i.e., neutron-neutron and proton-proton) and
 between the unlike nucleons (neutron-proton) play a decisive role in determining the isoscalar properties of nuclear matter.
In asymmetric nuclear matter, it is found that the isovector part of the s.p. potential
stems mainly from the $T=0$ channel, implying that the isospin-dependence
of the s.p. potential in asymmetric nuclear matter is governed by the interaction between
proton and neutron. We also checked that the contribution of the $T=0$ channel stems almost
completely from the contribution of the $T=0$ tensor $SD$ coupled channel, while
the contributions of the other isospin-singlet $T=0$ channels cancel almost completely.
On the one hand, the above result reveals an microscopic mechanism of the symmetry potential.
In asymmetric nuclear matter (for example, $\beta=0.6$), the neutron number is greater than the
proton number. As a result, a neutron may feel less interaction from surrounding protons and a proton
may feel stronger interaction from surrounding neutrons in asymmetric nuclear matter as compared with
the case of symmetric nuclear matter. Consequently, the effect of the symmetry
potential (i.e., the isovector part of the s.p. potential) is repulsive on neutrons and attractive on
protons due to the attraction of the tensor $SD$ coupled channel. On the other hand,
the present result on the s.p. potential is consistent with our previous conclusion for
the EOS of nuclear matter\cite{zuo:2009}. Therefore, we may conclude that both the isospin-dependence
of the s.p. potential and the EOS of asymmetric nuclear matter are determined mainly by the contribution
of the $T=0$ channel, implying a direct correspondence between the symmetry potential and symmetry energy.
In the transport models for heavy ion collisions, the direct input is the symmetry potential rather than the
symmetry energy. The present result confirms that one can use the transport model simulation to extract the
information of symmetry energy by comparing the isospin observables from experiments.

\section{Summary}
In the present paper, we have extended our previous work of Ref.\cite{zuo:2009} and
investigated the proton and neutron s.p. potentials
in isospin asymmetric nuclear matter by decomposing the potentials into
various spin-isospin $ST$ channels within the framework of the BHF approach extended to
include a microscopic three-body force. In symmetric nuclear matter, the s.p. potentials in both
isospin-singlet $T=0$ and isospin-triplet $T=1$ channels are attractive and they are comparable in magnitude.
In asymmetric nuclear matter, the isospin-dependence of the s.p. potentials in the $T=1$ channel turn out to be quite weak as compared with the s.p. potentials in the $T=0$ channel. As a consequence, the isovector part of the proton and neutron s.p. potentials is shown to be essentially  determined by the contribution from the $T=0$ channel in consistence with the conclusion obtained for the EOS of nuclear matter\cite{zuo:2009}.  The symmetry potential has also been decomposed into various spin-isospin channels. It is shown that the symmetry potential in the $T=0$ channel is much larger than that in the $T=1$ channel at not too high momenta, and the momentum dependence of the symmetry potential is governed by the contribution from the $T=0$ channel.

The present results are expected to provide some microscopic information for
constraining the isospin dependence of effective nucleon-nucleon
interactions in asymmetric nuclear medium.

\section*{Acknowledgments}
{The work was supported by the National Natural Science
Foundation of China (11175219, 10875151, 10740420550), the Major State Basic
Research Developing Program of China under No. 2007CB815004, the
Knowledge Innovation Project (KJCX2-EW-N01) of the Chinese Academy of
Sciences, the Chinese Academy of Sciences Visiting Professorship for Senior International
Scientists (Grant No.2009J2-26), and the CAS/SAFEA International Partnership Program for Creative
Research Teams (CXTD-J2005-1).}

\end{CJK*}
\end{document}